\documentclass[12pt]{article}
\usepackage{epsfig}
\usepackage{amssymb}
\usepackage{amsmath}
\usepackage{amsfonts}

\newcommand{\R}{\mathbb{R}}
\newcommand{\C}{\mathbb{C}}

\newcommand{\be}{\begin{equation}}
\newcommand{\ee}{\end{equation}}
\newcommand{\bea}{\begin{eqnarray}}
\newcommand{\eea}{\end{eqnarray}}

\newcommand{\kt}{\rangle}
\newcommand{\br}{\langle}

\newcommand{\ed}{\end{document}}

\begin{document}

\title{Krein-Space Formulation of ${\cal PT}$-Symmetry,
${\cal CPT}$-Inner Products, and\\ Pseudo-Hermiticity}
\author{\\
Ali Mostafazadeh
\\
\\
Department of Mathematics, Ko\c{c} University,\\
34450 Sariyer, Istanbul, Turkey\\ amostafazadeh@ku.edu.tr}
\date{ }
\maketitle

\begin{abstract}
Emphasizing the physical constraints on the formulation of a
quantum theory based on the standard measurement axiom and the
Schr\"odinger equation, we comment on some conceptual issues
arising in the formulation of ${\cal PT}$-symmetric quantum
mechanics. In particular, we elaborate on the requirements of the
boundedness of the metric operator and the diagonalizability of
the Hamiltonian. We also provide an accessible account of a
Krein-space derivation of the ${\cal CPT}$-inner product that was
widely known to mathematicians since 1950's. We show how this
derivation is linked with the pseudo-Hermitian formulation of
${\cal PT}$-symmetric quantum mechanics.

\vspace{5mm}

\vspace{2mm}


\end{abstract}

%\tableofcontents
%\textheight = 22cm \topskip = -1cm \topmargin = -1cm

\section{Introduction}

``In theoretical physics, there are a number of ideas  that are
periodically (re)discovered and then forgotten.'' This was pointed
out to me by George Sudarshan when I was a graduate student. By
now, I have encountered a few concrete examples of these
rediscoveries. I have also realized that sometimes they undergo
mutations while they are repeated/reproduced and follow an
evolutionary pattern in time. Occasionally they lead to an
improved version of a previously known but forgotten idea that has
a wider domain of application. Such rediscoveries are by no means
inferior to genuine discoveries, for they may be useful in dealing
with certain problems that their previous versions could not
handle.

I have elaborated on a recent example of such a ``rediscovery'' in
\cite{cjp-2003}. It has to do with a recent version of the old
notion of ``a pseudo-Hermitian operator'' used in the context of
indefinite-metric quantum theories \cite{indefinite-phys} --
\cite{nakanishi} and indefinite-metric linear spaces
\cite{indefinite-math,bognar,azizov}, namely the one proposed in
\cite{p1,p23}. The main purpose for introducing this notion was to
understand the spectral properties of a class of ${\cal
PT}$-symmetric Hamiltonians \cite{bender-prl-98}. The ensuing
developments have not only achieved this purpose and led to a
consistent formulation of unitary quantum systems based on a
general class of non-Hermitian\footnote{We will follow the
terminology used in the subject, and call a linear operator
``Hermitian'' if it is represented by a Hermitian matrix in a
standard basis of the vector space it acts in. Below, we will give
a precise definition.} Hamiltonians
\cite{jpa-2004b,jpa-2005a,other-models,p67}, but perhaps more
importantly played a central role in obtaining a complete solution
of one of the oldest problems of modern theoretical physics,
namely the problem of devising a genuine quantum mechanical
treatment of Klein-Gordon and similar fields \cite{KG}.

The key ingredient of the new notion of a pseudo-Hermitian
operator is that it does not rely on a fixed pseudo-metric. Indeed
the treatment of these operators given in \cite{p1,p23} involves
constructing all possible compatible pseudo-metric operators that
are associated with a pseudo-Hermitian operator $H$. In
particular, whenever possible it yields the form of the associated
positive-definite metric operators and the corresponding inner
products that render $H$ self-adjoint \cite{p23,p4}.

In \cite{bbj}, the authors propose an alternative construction for
a positive-definite inner product for a class of ${\cal
PT}$-symmetric Hamiltonians. Although the point of departure and
the approach leading to the inner product of \cite{bbj}, i.e., the
so-called ${\cal CPT}$-inner product, are different from those of
\cite{p23}, as shown in \cite{jmp-2003,jpa-2005a} this inner
product belongs to a special class of those constructed in
\cite{p23}. See also \cite{p4}.\footnote{A concrete demonstration
of the fact that there are physically acceptable inner products
that do not belong to the set of ${\cal CPT}$-inner products is
given in \cite{p67}.}

The purpose of this paper is two fold: Firstly, I wish to draw
attention to the physical constraints that need be imposed on any
mathematical framework that aims at describing a unitary ${\cal
PT}$-symmetric or pseudo-Hermitian quantum system. Secondly, I
wish to show, using a minimal amount of mathematical formalism,
that indeed the discovery of the ${\cal CPT}$-inner product
\cite{bbj} is another example of the ``rediscoveries'' I eluded to
above. As far as I could trace, its original form appeared in a
series of papers \cite{nevanlinna} by Rolf Nevanlinna published
between 1952 and 1956. Nevanlinna's construction is reviewed in
mathematics \cite{bognar,azizov} as well as physics
\cite{nagy,nakanishi} literature. It was indeed discussed by
Japaridze \cite{japaridze} in 2001 in the context of ${\cal
PT}$-symmetry, rediscovered by Bender, Brody, and Jones \cite{bbj}
in 2002, employed by Albeverio and Kuzhel \cite{albaverio} in
2004, and more recently by Tanaka \cite{tanaka,tanaka2} in 2006.

In what follows, I will first outline the physical constraints
that limit the vast and virtually unbounded arena of mathematical
possibilities for the problem and then give an accessible
derivation of Nevanlinna's positive-definite inner product. This
will in turn provide an opportunity to discuss various aspects of
the problem, e.g., the connection to the pseudo-Hermitian
formulation, the justification for the diagonalizability of the
Hamiltonian, and the issue of the boundedness of the
(pseudo-)metric operator.

\section{Must Quantum Observables Be Hermitian?}

In 1918, when Einstein received Weyl's letter asking his opinion
on the latter's paper on gauge theories, Einstein's response was
the following \cite{enstein}: ``Except for the agreement with
reality, it is in any case a grand intellectual achievement.''
Einstein's reaction to Kaluza's 5-dimensional unification of
gravity and electrodynamics was not different \cite{enstein2}.
These are two instructive examples of how demanding physical
considerations can become when mathematics is used to attack a
particular physical problem. The same is true about quantum
mechanics. It is not a lucky accident that out of the infinity of
possibilities for function spaces that mathematicians are so eager
to use in dealing with various problems in applied mathematics, it
is only the (separable) Hilbert space that was found suitable by
von~Neumann to formulate quantum mechanics \cite{von-neumann}.
This was dictated by the physical constraints. Today, the same
highly restrictive physical constraints persist in formulating a
quantum theory based on a non-Hermitian Hamiltonian. Therefore, it
seems necessary to seek for appropriate mathematical models only
from among those that are compatible with the physical
constraints.

A central objective of the recent study of ${\cal PT}$-symmetric
Hamiltonians is to devise a quantum theory in which the dynamics
is generated by the usual time-dependent Schr\"odinger equation,
    \be
    i\hbar\frac{\partial}{\partial t}\Psi(z,t)=  H\Psi(z,t),
    ~~~~~t\in\R,~~~z\in\C,
    \label{sch-eq}
    \ee
where the Hamiltonian operator $H$ is a Schr\"odinger operator,
    \be
      H=-\frac{d^2}{dz^2}+v(z),
    \label{sch-op}
    \ee
with a complex-valued potential $v:\C\to\C$, such as
\cite{bender-prl-98}
    \be
    v(z)=-(iz)^N,~~~~N\in\R^+.
    \label{v=}
    \ee
The state vectors $\Psi$ of the theory belong to a function space
${\cal V}_{_\Gamma}$ that, as we explain below, is defined using a
certain contour $\Gamma$ in complex plane $\C$. Specifically,
$\Gamma$ is obtained from the real axis $\R$ by an invertible
deformation. As is common practice, we shall often identify
$\Gamma$ and a parameterized (piecewise regular) curve
$\gamma:\R\to\C$ such that $\Gamma=\{\gamma(x)|x\in\R\}$. The most
convenient $\gamma$ is the one corresponding to the arc-length
parametrization of $\Gamma$. We will assume that
$\lim_{x\to\pm\infty}\gamma(x)=\infty$, where the second $\infty$
stands for the point at infinity, and that $\Gamma$ is symmetric
about imaginary axis whenever $v$ is ${\cal PT}$-invariant, i.e.,
$v(r)^*=v(-r)$ for $r\in\R$, \cite{jpa-2005a}.

The function space ${\cal V}{_\Gamma}$ has the following
properties.
\begin{itemize}
\item[](1) It is a complex vector space;

\item[](2) It includes the domain of $  H$ at least for all
potentials $v$ for which the eigenvalue problem
    \be
      H \Psi_n(z)=E_n\Psi_n(z),~~~~~~z\in\Gamma,
    \label{sch-eq-t-indep}
    \ee
viewed as a holomorphic differential equation and solved along
$\Gamma$ is well-posed \cite{Hille}. The latter condition is
fulfilled provided that one imposes a set of appropriate boundary
conditions at the infinities $\gamma(\pm\infty)$ of $\Gamma$. For
typical polynomial potentials considered in the literature
\cite{polynomial,shin}, one requires a sufficiently rapid,
typically exponential, decay at the infinities:
    \be
    |x|\to\infty~~~{\rm implies}~~~
    |\Psi(\gamma(x))| \to 0 ~\mbox{exponentially;}
    \label{exp}
    \ee
\item[](3) It is endowed with an appropriate complete metric such
that the domain of $  H$ for all potentials with the properties
mentioned in (2) is dense in ${\cal V}_{_\Gamma}$. This is usually
achieved by selecting a norm, i.e., promoting ${\cal V}_{_\Gamma}$
to a Banach space. In particular, given the boundary
condition~(\ref{exp}), one can view the state vectors $\Psi$ as
elements of the separable Hilbert space $L^2(\Gamma)$, i.e., the
space of square-integrable functions $\Psi:\Gamma\to\C$,
\cite{dorey,jpa-2005a}, where $\Gamma$ is viewed as a
one-dimensional submanifold of $\C=\R^2$ with the metric (and
hence integral measure) induced from the Euclidean metric on
$\C=\R^2$, \cite{cecile}. If one adopts the arc-length
parametrization $\gamma(x)$ of $\Gamma$, the contribution to the
induced integral measure is identically 1, and the $L^2$-inner
product along $\Gamma$ has the form
    \be
    \br\Psi|\Phi\kt_{_\Gamma}:=\int_{\Gamma}
    \Psi(z)^*\Phi(z)dz=\int_{\R}\Psi(\gamma(x))^*\Phi(\gamma(x))dx.
    \label{L2}
    \ee
\end{itemize}

The above discussion suggests that the state space ${\cal
V}_{_\Gamma}$ is to be identified as a (topological) vector space
with the Hilbert space
$$L^2(\Gamma):=\{\Psi:\Gamma\to\C~|~\br\Psi|\Psi\kt_{_\Gamma}<0\}.$$
This is a consequence of purely mathematical considerations. The
state space ${\cal V}_{_\Gamma}$ is furthermore required to be a
separable Hilbert space with an inner product that renders the
Hamiltonian operator $  H$ self-adjoint. As we show in the
remainder of this section this is a direct and unavoidable
consequence of the measurement axiom which we intend to adopt.
Specifically, we assume the following.
\begin{itemize}

\item[](i) There is a positive-definite inner product
$\br\cdot,\cdot\kt_+:{\cal V}_{_\Gamma}^2\to\C$ on ${\cal
V}_{_\Gamma}$ such that ${\cal H}_{\rm phys}:=({\cal V}_{_\Gamma},
\br\cdot,\cdot\kt_+)$ is a separable Hilbert space.

\item[](ii) The (pure) states of the system are in one-to-one
correspondence with the rays (one-dimensional subspaces) of ${\cal
H}_{\rm Phys}$ each of which is described by a nonzero state
vector $\Psi\in{\cal H}_{\rm Phys}$.

\item[](iii) The observables $O$ of the theory are linear
operators acting in ${\cal H}_{\rm phys}$.

\item[](iv) Suppose that $O$ is an observable having a discrete
spectrum\footnote{The case of continuum spectrum can be treated
similarly \cite{von-neumann}.} and that the system is in the state
described by the state vector $\Psi$. Upon measuring $O$, the
state of the system collapses onto a state described by a state
vector $\Omega_n$. The measuring device records $\omega_n$ if and
only if $\Omega_n$ is an eigenvector of $O$ with eigenvalue
$\omega_n$. The probability of measuring $\omega_n$ is given by
    \be
    {\rm Prob}_n(\Psi):=
    \frac{\br\Lambda_n\Psi,\Lambda_n\Psi\kt_+}{\br\Psi,\Psi\kt_+},
    \label{prob}
    \ee
where $\Lambda_n$ is the projection operator\footnote{It is a
linear operator acting in ${\cal V}_{_\Gamma}$ and satisfying $
\Lambda_n^2=\Lambda_n$.} onto the subspace of ${\cal V}_{_\Gamma}$
consisting of the eigenvectors of $O$ with eigenvalue $\omega_n$.

\item[](v) The probability ${\rm Prob}_n(\Psi)$ of measuring
$\omega_n$ is a continuous function of time, i.e., if one measures
$\omega_n$ at time $t=0$ and repeats the same measurement at time
$t>0$, the probability of measuring a different eigenvalue than
$\omega_n$ tends to zero as $t\to 0$.

\end{itemize}

Because measuring devices read real numbers, the spectrum of an
observable is necessarily real. Furthermore, the following two
physical requirements imply that the set ${\rm Span}(O)$ of linear
combinations of the eigenvectors of $O$ must be dense in ${\cal
H}_{\rm Phys}$.
    \begin{itemize}
    \item[](a) There is a dense set ${\cal D}$ of elements $\Psi$ of
${\cal H}_{\rm Phys}$ that describe states of the system in which
$O$ can be measured\footnote{This means that every nonzero state
vector $\Psi\in{\cal H}_{\rm Phys}$ describes a physically
accessible state of the system; it can be approximated with any
desired accuracy by the elements of ${\cal D}$.};
    \item[](b) The
probability of all possible outcomes of a measurement must add up
to unity.
    \end{itemize}
Note also that (iv) puts an strong restriction on the projection
operators, namely that $\Lambda_m\Lambda_n= \delta_{mn}\Lambda_n$.
This implies the existence of an orthogonal basis of ${\cal
H}_{\rm Phys}$ consisting of the eigenvectors of $O$.\footnote{The
generalization of the situation for the cases that the spectrum of
$O$ is not discrete leads to the requirement that $O$ has a
spectral resolution of the identity \cite{yosida}.} This together
with the reality of the spectrum of $O$ yield the condition:
``\emph{$O$ must be a densely defined self-adjoint operator acting
in ${\cal H}_{\rm Phys}$}.''

It should be emphasized that \emph{although the measurement axiom
does not fix or restrict the choice of the defining inner product
of the physical Hilbert space, it makes the self-adjointness of
the observables absolutely necessary.} The condition of the
Hermiticity of the observables $O$ and in particular the
Hamiltonian that is adopted in the conventional quantum mechanics
is therefore a valid and indisputable condition provided that it
is interpreted as self-adjointness of $O$ as an operator acting in
the physical Hilbert space ${\cal H}_{\rm Phys}$, i.e., for every
$\Psi$ and $\Phi$ in the domain of $O$,
    \be
    \br\Psi,O\Phi\kt_+=\br O\Psi,\Phi\kt_+.
    \label{self-adjoint}
    \ee
This may be called \emph{Hermiticity with respect to the inner
product $\br\cdot,\cdot\kt_+$}. This is not the same notion of
\emph{Hermiticity} that is used in the literature in ${\cal
PT}$-symmetry. The latter is defined in terms of a pre-assigned
\emph{reference inner product} $\br\cdot|\cdot\kt$ that is
generically different from $\br\cdot,\cdot\kt_+$ and hence
physically inadmissible. The typical example of the reference
inner product $\br\cdot|\cdot\kt$ is the $L^2$-inner product
$\br\cdot|\cdot\kt_{_\Gamma}$ of Eq.~(\ref{L2}). Therefore the
answer to the question posed in the title of this section and that
of \cite{bender-ajp} is that it depends on the definition of the
term ``Hermitian.'' If one defines it with respect to the
$L^2$-inner product which is equivalent to saying that $O$ is
Hermitian if it can be expressed as the integral operator,
    \[(O\Psi)(\gamma(x))=
    \int_{\R}O(x,y)\Psi(\gamma(y))dy,\]
with a kernel satisfying $O(x,y)^*=O(y,x)$ for all $x,y\in\R$,
then the answer is NO. But this answer must be qualified by saying
that $O$ must nevertheless be Hermitian with respect to some
positive-definite inner product $\br\cdot,\cdot\kt_+$. The issue
of the existence and construction of $\br\cdot,\cdot\kt_+$ is
currently a subject of active research
\cite{p23,bbj,bbj-prd,other-models,jones-2005,p67,scholtz,znojil,p69,fring}.
For an earlier investigation see \cite{quasi}.

\section{The Krein-Space or Indefinite-Metric Formulation}

Let ${\cal V}$ be a complex vector space. Then a function $Q:{\cal
V}^2\to\C$ with domain ${\cal V}^2$ is said to be a \cite{kato}
    \begin{itemize}
    \item {\em nondegenerate form}, if ``for all $v\in{\cal V}$,
    $Q(v,w)=0$'' implies ``$w=0$'';
    \item {\em Hermitian form}, if $Q(v,w)^*=Q(w,v)$ for all
    $v,w\in{\cal V}$;
    \item {\em sesquilinear form}, if
    $Q(u,\alpha v+\beta w)=
    \alpha Q(u,v)+\beta Q(u,w)$ and
    $Q(\alpha u+\beta v,w)=\alpha^*Q(u,w)+\beta^* Q(v,w)$,
    for all
    $u,v,w\in{\cal V}$ and $\alpha,\beta\in\C$;
    \item {\em positive-definite form}, if $Q(v,v)\in\R^+$ for all
    $v\in{\cal V}-\{0\}$.
    \end{itemize}
Furthermore, if ${\cal V}$ is endowed with a norm \cite{kato}
denoted by  $\parallel\cdot\parallel$, then $Q$ is called a
    \begin{itemize}
\item {\em bounded form}, if there exists $c\in\R^+$ such that
    $|Q(v,w)|\leq c \parallel v\parallel\,\parallel w\parallel$
    for all $v,w\in{\cal V}$.
    \end{itemize}
A non-degenerate Hermitian sesquilinear form is called a
\emph{pseudo-inner product}. A positive-definite pseudo-inner
product is called a \emph{positive-definite inner product} or
simply an \emph{inner product}. A pseudo-inner product that is not
positive-definite is called an \emph{indefinite inner product}.

If one endows ${\cal V}$ both with a norm
$\parallel\cdot\parallel$ and a bounded positive-definite inner
product $\br\cdot,\cdot\kt:=Q(\cdot,\cdot)$, then the norm $\sqrt
{\br\cdot,\cdot\kt}$ associated with $\br\cdot,\cdot\kt$ defines a
topology on ${\cal V}$ which is identical with the topology
defined by the defining norm
$\parallel\cdot\parallel$.\footnote{This means that both norms
define the same notion of the convergence for sequences in ${\cal
V}$.}

Suppose that $\br\cdot,\cdot\kt$ is a positive-definite inner
product on ${\cal V}$ and $\parallel\cdot\parallel:=\sqrt
{\br\cdot,\cdot\kt}$ is the corresponding norm. It is not
difficult to show that there is a one-to-one correspondence, given
by
    \be
    Q_T(v,w)=\br v,Tw\kt,
    \label{form}
    \ee
between bounded sesquilinear forms $Q_T$ on ${\cal V}$ and
everywhere-defined bounded linear operators $T:{\cal V}\to{\cal
V}$, \cite{kato}. If $Q_T$ is non-degenerate (respectively,
Hermitian or positive-definite), $T$ is invertible (respectively,
self-adjoint or positive-definite). Here invertible means that $T$
is one-to-one, onto, and $T^{-1}$ is bounded \cite{kato},
self-adjoint means: $\br v,Tw\kt=\br Tv,w\kt$ for all $v,w\in{\cal
V}$, and positive-definite means: $\br v,Tv\kt\in\R^+$ for all
$v\in{\cal V}-\{0\}$.

A \emph{Krein Space} \cite{azizov} is a complex vector space
${\cal V}$ that is endowed with an indefinite inner product
$(\cdot,\cdot)$ and has a pair of vector subspaces ${\cal V}_\pm$
with the following properties.
    \begin{itemize}
\item $\pm (v_\pm,v_\pm)>0$ for all $v_\pm\in{\cal V}_\pm-\{0\}$;

\item ${\cal V}_-$ and ${\cal V}_+$ are orthogonal,
i.e.,$(v_+,v_-)=0$ for all $v_\pm\in{\cal V}_\pm-\{0\}$;

\item ${\cal V}$ is a direct sum of ${\cal V}_-$ and ${\cal V}_+$,
    \be
    {\cal V}={\cal V}_-\oplus{\cal V}_+,
    \label{direct}
    \ee
i.e., for all $v\in{\cal V}$ there are unique $v_\pm\in{\cal
V}_\pm$ such that $v=v_-+v_+$;

\item ${\cal V}_\pm$ endowed with the positive-definite inner
products obtained by restricting $\pm(\cdot,\cdot)$ onto ${\cal
V}_\pm$ are separable Hilbert spaces.
\end{itemize}

A principal result of the theory of Krein spaces is that a Krein
space may be endowed with a positive-definite inner product
$\br\cdot,\cdot\kt_+:{\cal V}^2\to\C$ and turned into a separable
Hilbert space. This is achieved using the projection operators
$\Pi_\pm:{\cal V}\to{\cal V}$ associated with the orthogonal
direct sum decomposition (\ref{direct}); $\Pi_\pm$ projects every
$v\in{\cal V}$ onto its component $v_\pm$ in ${\cal V}_\pm$.
Clearly, $\Pi_-=I-\Pi_+$, where $I$ denotes the identity operator
acting in ${\cal V}$. The inner product $\br\cdot,\cdot\kt_+$ has
the form \cite{nevanlinna,azizov,nagy,langer}:
    \be
    \br v,w\kt_+:=(\Pi_+ v,\Pi_+ w)-(\Pi_- v,\Pi_- w)=
    (v,Cw),
    \label{C}
    \ee
where
    \be
    C:=\Pi_+-\Pi_-=2\Pi_+-I.
    \label{C=}
    \ee
The operator $C$ is actually a \emph{grading operator}, for
    \be
    {\cal V}_\pm=\{v\in{\cal V}~|~Cv=\pm v\}.
    \label{Vpm}
    \ee
In particular it is an involution: $C^2=I$. This operator which in
the mathematical literature is usually denoted by $J$ and referred
to as the ``\emph{fundamental symmetry operator}'' \cite{langer}
is precisely the ``charge-conjugation operator'' of Bender, Brody,
and Jones \cite{bbj} whenever the Krein-space theory is applied in
the study of ${\cal PT}$-symmetric Hamiltonians
\cite{japaridze,albaverio}.

Because ${\cal V}$ endowed with $\br\cdot,\cdot\kt_+$ is a
separable Hilbert space that admits the direct sum
decomposition~(\ref{direct}), there is an orthonormal basis
$\{\xi^\pm_n\}$ of this Hilbert space such that $\xi^\pm_n\in{\cal
V}_\pm$ for all $n=0,1,2,\cdots$. We can relabel the basis vectors
according to
    \be
    \xi^\pm_n\to \psi_n:=\left\{\begin{array}{ccc}
    \xi^+_{\frac{n}{2}}&{\rm for}& \mbox{$n$ is even}\\
    \xi^-_{\frac{n-1}{2}} &{\rm for}& \mbox{$n$ is odd}.
    \end{array}\right.
    \label{f=}
    \ee
This yields
    \be
    \br\psi_m,\psi_n\kt_+=\delta_{mn},~~~~~~~~~~m,n=0,1,2,\cdots,
    \label{ortho-psi}
    \ee
and, in view of (\ref{C}) and (\ref{Vpm}) -- (\ref{ortho-psi}),
    \be
    (\psi_m,\psi_n)=(-1)^m\delta_{mn},~~~~~~~~~~m,n=0,1,2,\cdots.
    \label{basis}
    \ee
Furthermore, we have
    \be
    C=\sum_n (-1)^n \Lambda_n,
    \label{C=exp}
    \ee
where $\Lambda_n$ denotes the projection operator defined by
    \be
    \Lambda_n w:=\br \psi_n,w\kt_+\psi_n,~~~~~~~\mbox{for all}~~~~~
    w\in{\cal V}.
    \label{Lambda=}
    \ee

Next, consider the case that ${\cal V}$ has the structure of a
separable Hilbert space with another inner product
$\br\cdot|\cdot\kt:{\cal V}^2\to\C$ such that the corresponding
norm $\parallel\cdot\parallel$ defines the same topology as the
norm $\parallel\cdot\parallel_+$ associated with the inner product
$\br\cdot,\cdot\kt_+$. Let us denote the Hilbert space $({\cal V},
\br\cdot|\cdot\kt)$ by ${\cal H}$ and the Hilbert space $({\cal
V}, \br\cdot,\cdot\kt_+)$ by ${\cal H}_+$. Then, as we mentioned
above, there is an everywhere-defined, bounded, positive-definite,
invertible operator $\eta_+:{\cal H}\to{\cal H}$ such that
\cite{kato}
    \be
    \br\cdot,\cdot\kt_+=\br\cdot|\eta_+\cdot\kt.
    \label{inn}
    \ee
Similarly we have an everywhere-defined, bounded, self-adjoint,
invertible operator $P:{\cal H}\to{\cal H}$ such that\footnote{As
I explain in \cite{p74}, an everywhere-defined self-adjoint
operator is necessarily bounded and the ontoness of $P$ ensures
that its inverse is bounded. Hence the conditions listed here may
be summarized as: ``$P$ is a Hermitian (self-adjoint)
automorphism,'' \cite{p1}.}
    \be
    (\cdot,\cdot)=\br\cdot|P\cdot\kt.
    \label{Parity}
    \ee
Combining (\ref{C}), (\ref{inn}), and (\ref{Parity}), we have
\cite{jmp-2003,jpa-2005a}
    \be
    C=C^{-1}=\eta_+^{-1}P.
    \label{C-eta}
    \ee

Because the span of the vectors $\psi_n$ is a dense subset of
${\cal H}$, we can identify $\{\psi_n\}$ with a basis of ${\cal
H}$. It is important to note that in general this basis fails to
be orthogonal.

Next, observe that the unique positive square root
$\rho=\sqrt{\eta_+}$ of $\eta_+$ viewed as an operator mapping
${\cal H}_+$ onto ${\cal H}$ satisfies
    \be
    \br\rho v|\rho w\kt=\br v|\eta_+ w\kt=\br v,w\kt_+,~~~~~
    \mbox{for all}~~~~~~v,w\in{\cal H}_+.
    \label{unitary}
    \ee
Hence $\rho:{\cal H}_+\to{\cal H}$ is a unitary operator
\cite{jpa-2003}. A simple consequence of (\ref{unitary}) is that
$\zeta_n:=\rho \psi_n$ form an orthonormal basis of ${\cal H}$.
Because $\eta_+$ is an everywhere-defined bounded invertible
operator, so is its square root $\rho$. This in turn means that
$\{\psi_n\}$ is related to an orthonormal basis ($\{\zeta_n\}$)
via a bounded invertible operator. Such a basis is called a
\emph{Riesz basis}, \cite{goh-ker}. A characteristic property of
every Riesz basis is that it can be extended to a biorthonormal
system \cite{goh-ker}, i.e., there is a (Riesz) basis $\{\phi_n\}$
of ${\cal H}$, such that
    \bea
    &&\br\phi_m|\psi_n\kt=\delta_{mn},
    \label{ortho}\\
    &&\sum_n \br \phi_n|w\kt\psi_n= \sum_n \br \psi_n|w\kt\phi_n=w
    ~~~~{\rm for~all}~~~w\in{\cal H}.
    \label{biortho}
    \eea
Rewriting the latter relation using the Dirac notation one finds
the familiar condition \cite{p1,p23}:
    \be
    \sum_n |\psi_n\kt\br\phi_n|=I=\sum_n |\phi_n\kt\br\psi_n|.
    \label{biortho-dirac}
    \ee

In view of (\ref{ortho-psi}), (\ref{inn}), and (\ref{ortho}), we
can relate $\phi_n$ to $\psi_n$ via
    \be
    \phi_n=\eta_+\psi_n.
    \label{phi=}
    \ee
Inserting this equation in (\ref{Lambda=}) and using (\ref{inn})
and the self-adjointness of $\eta_+$, we then find
    \[\Lambda_n\psi=\br\psi_n|\psi\kt_+\psi_n=
    \br\psi_n|\eta_+\psi\kt\psi_n=\br\eta_+\psi_n|\psi\kt\psi_n=
    \br\phi_n|\psi\kt\psi_n,
    ~~~~~\mbox{for all}~~~~~\psi\in{\cal H}.\]
We can express this result using the Dirac notation in the form:
    \be
    \Lambda=|\psi_n\kt\br\phi_n|.
    \label{Lambda=2}
    \ee
Combining this relation with (\ref{C=exp}), we have
    \be
    C=\sum_n (-1)^n |\psi_n\kt\br\phi_n|.
    \label{C=3}
    \ee
This equation was initially derived in \cite{jmp-2003} in an
attempt to show how the results of \cite{bbj} relate to those of
\cite{p23}.

Next, we use (\ref{phi=}), (\ref{biortho-dirac}), (\ref{C-eta}),
and (\ref{C=3}) to establish
    \bea
    \eta_+&=&\sum_n |\phi_n\kt\br\phi_n|,
    \label{eta=}\\
    \eta_+^{-1}&=&\sum_n |\psi_n\kt\br\psi_n|,
    \label{eta-inv=}\\
    P&=&\sum_n (-1)^n |\phi_n\kt\br\phi_n|.
    \label{P=}
    \eea
These are among the basic equations of the pseudo-Hermitian
treatment of ${\cal PT}$-symmetric Hamiltonians developed in
\cite{p1,p23,p4,jmp-2003}. It is absolutely essential to realize
that the analysis leading to (\ref{C=3}) -- (\ref{P=}) applies in
any Krein-space setting.

\section{Boundedness of the Metric and Diagonalizability
 of the Hamiltonian}

In applying the properties of pseudo-Hermitian operators \cite{p1}
to ${\cal PT}$-symmetric Hamiltonians $H$ that have a real and
discrete spectrum, one adopts a separable (reference) Hilbert
space ${\cal H}$ containing the domain of $H$ as a dense subset
(so that the adjoint $H^\dagger$ of $H$ is well-defined
\cite{reed-simon}) and assumes that as an operator acting in
${\cal H}$ the Hamiltonian $H$ is diagonalizable
\cite{p23,p4,jpa-2004b}. This precisely means that there is a
Riesz basis $\{\psi_n\}$ of ${\cal H}$ consisting of the
eigenvectors of $H$.\footnote{A direct consequence of the
identification of $\psi_n$ with the eigenvectors of $H$ is
$[C,H]=0$.} The fact that as a vector space the Hilbert space must
only include the closed span of the eigenvectors of $H$ is a
physical necessity.\footnote{Otherwise, there will be states for
which $H$ cannot be measured!} Therefore, the existence of a basis
consisting of the eigenvectors of $H$ is easy to justify. The
additional condition that this basis be a Riesz basis is a
mathematical requirement. One can enforce it by choosing an
appropriate reference Hilbert space ${\cal H}$, i.e., an
appropriate inner product $\br\cdot|\cdot\kt$ on the vector space
of state vectors that renders the latter a separable Hilbert
space. The condition that such an inner product must exist is a
physical requirement on $H$. But the choice of $\br\cdot|\cdot\kt$
is not dictated by physics and can be made arbitrarily. For
example one may define the Hilbert space of the theory by taking
the span of the eigenvectors of $H$, endow it with the inner
product that renders the eigenvectors orthonormal, and complete
the resulting inner product space into a Hilbert space as proposed
by Kretschmer and Szymanowski~\cite{k-s}.

The approach of \cite{k-s} directly gives the physical Hilbert
space of the system, but in essence uses a representation of the
system that is based on the Hilbert space $\ell_2$ of square
summable complex sequences \cite{cjp-2004b}. In this
representation the Hamiltonian enters only in the form of its
eigenvalues. Furthermore, in view of the form of the inner product
on $\ell_2$:
    \[\br \{s_n\},\{t_n\}\kt:=\sum_{n=0}^\infty s_n^*t_n,\]
the calculation of physical quantities, such as the expectation
value of observables, involves summing complicated series. To
avoid dealing with such an important practical problem, one maps
$\ell_2$ onto $L_2$ by an appropriate unitary operator and
performs the calculations in this representation that uses $L_2$
as the physical Hilbert space of the system. Alternatively, one
may perform a unitary mapping of the physical Hilbert space onto
$L_2$ directly to arrive at this $L^2$-representation \cite{k-s}.
The fact that this can always be done is a well-known consequence
of the uniqueness theorem for infinite-dimensional separable
Hilbert spaces \cite{cjp-2004b}. But there is no unique or
systematic (canonical) way of choosing this unitary mapping and
the corresponding alternative $L^2$-representation. One is
naturally inclined to determine and make a choice in which the
Hamiltonian takes a simpler form in the $L^2$-representation. How
one can achieve this, however, is not known to the present author.

The main advantage of the method of \cite{k-s} is that it avoids
the technical problems related to the boundedness of the metric
operators. As we mentioned above the metric operators $\eta_+$
that define (topologically equivalent) positive-definite inner
products are necessarily everywhere-defined, bounded, and
invertible. These conditions are sometimes ignored in constructing
concrete examples of a metric operator for models typically
defined in ${\cal H}=L^2(\Gamma)$. These constructions involve
obtaining an operator $\eta_+$ that satisfies the
pseudo-Hermiticity condition
    \be
    H^\dagger=\eta_+ H\eta_+^{-1},
    \label{ph}
    \ee
is formally Hermitian, i.e., there is a dense subset ${\cal
S}\subseteq{\cal H}$ in which $\eta_+^\dagger=\eta_+$, and
positive-definite, i.e., $\eta_+$ has a real positive-definite
spectrum. Such an operator defines an inner product on ${\cal S}$.
The resulting inner product is generally incomplete. It can be
completed to a separable Hilbert space. But the resulting space
generally differs from ${\cal H}$, for they have different
topologies.\footnote{Such non-invertible or unbounded ``metric
operators'' have an important application \cite{resonances1} in
the context of the method of complex scaling in the study of
resonances \cite{resonances2}.} This observation is sometimes
viewed as a shortcoming of the pseudo-Hermitian treatment of
${\cal PT}$-symmetric systems \cite{k-s}. This view seems to
ignore the fact that the failure to construct examples of metric
operators that satisfy all the above-mentioned conditions cannot
be accepted as a proof of their nonexistence.

The pseudo-Hermitian treatment of the ${\cal PT}$-symmetric models
considered in \cite{p1,p23,jpa-2005a} is based on choosing
$L^2(\Gamma)$ as the reference Hilbert space ${\cal H}$. This has
the advantage of making the relationship between the usual
Hermitian quantum mechanics, where ${\cal H}$ is the physical
Hilbert space, more transparent. Indeed, this approach is the only
known one that allows for the determination of an underlying
classical system (with the correct number of degrees of freedom
\cite{pla-2006}) and a consistent quantization scheme that links
the latter to the original quantum system,
\cite{jpa-2004b,other-models,p67}. These are absolutely essential
for assigning physical meaning to the operators that one
identifies with the observables of the theory. The only
disadvantage of this approach is the possibility that $H$ may turn
out to be non-diagonalizable as an operator acting in
$L^2(\Gamma)$, i.e., there exists a set of Hilbert spaces ${\cal
H}$ in which $H$ acts as a diagonalizable operator but
$L^2(\Gamma)$ does not belong to this set.

Contrary to the claims made in \cite{tanaka2}, the requirement of
the existence of a biorthonormal system consisting of eigenvectors
of the Hamiltonian operator and its adjoint is well-justified by
physical considerations. According to quantum measurement theory,
one cannot call a (differential) operator $H$ ``the Hamiltonian of
a unitary quantum system,'' if the set of its eigenvectors is
incomplete or more generally if there does not exist a separable
Hilbert space in which $H$ acts as a densely defined self-adjoint
operator.\footnote{A class of examples of such operators are those
involving spectral singularities \cite{samsonov,curtright}.}

In practice, given a differential operator $H$ with a well-posed
eigenvalue problem, one has access to its domain as a function
space. To embed this space (as a closed subspace) in a reference
Hilbert space ${\cal H}$ such that the set of eigenvectors of $H$
form a Riesz basis of ${\cal H}$ is a mathematical problem that
must be considered separately from the eigenvalue problem for $H$.
If such a Hilbert space does not exist, $H$ is not a viable
candidate for the Hamiltonian operator of a physical system. If
such a Hilbert space exists then the eigenvectors of $H$ form a
Riesz basis and they can be extended to a biorthonormal system,
i.e., as an operator acting in ${\cal H}$, $H$ is diagonalizable.
Determining the class of differential operators fulfilling this
condition is an interesting and admittedly difficult mathematical
problem.

The physical problem may be devised and investigated only after
one is provided with a differential operator $H$ and a separable
Hilbert space ${\cal H}$ in which $H$ is diagonalizable. Clearly
there are an infinity of (unitarily equivalent) separable Hilbert
spaces ${\cal H}$ rendering such a differentiable operator
diagonalizable. All these choices are clearly physically
equivalent \cite{cjp-2004b} and for all of them the construction
outlined in \cite{p23} for a physically admissible
positive-definite inner product applies. The question whether
$L^2(\Gamma)$ is among the possible choices for ${\cal H}$ must be
addressed for each $H$ separately. Based on numerical evidence it
is assumed that this is the case for most ${\cal PT}$-symmetric
potentials considered in the literature \cite{bbj}.

\section{Concluding Remarks}

The postulate that physical states of a quantum system are
represented by rays in a separable Hilbert space and the quantum
measurement postulate are as indispensable for any quantum theory,
as is the Schr\"odinger time evolution. This is independent of
whether one begins with a mathematical formalism that does not
involve a Hilbert space structure from the outset. A serious
attempt to formulate a genuine generalization of (or an
alternative to) quantum mechanics must begin with a clear and
comprehensive description of its postulates. This must follow with
a discussion of interpretational issues. A mathematical scheme for
achieving such a theory, however rigorous it may be, is to be
rejected, if it does not obey the conditions set by the physical
reality. The history of modern physics shows that the main
guidelines for real progress are physical considerations.

Imposition of the demands of physics upon mathematical models is
also a delicate issue. One must be extremely careful in ascribing
the property of being \emph{physical} to a mathematical construct.
For example, some of the workers on the subject use the following
argument as a motivation for their investigations: ``${\cal
PT}$-symmetry is a more physical condition than Hermiticity,
because it means space-time reflection symmetry.'' This would be
true, if the parity operator ${\cal P}$ described
``space-reflection.'' This happens if and only if the space, i.e.,
the position observable of a particle, is identified with the
usual $x$ operator. The latter is well-known not to be an
observable in almost all known ${\cal PT}$-symmetric models. For
these models, therefore, ${\cal PT}$ does not describe a
space-time reflection, and \emph{${\cal PT}$-symmetry is not a
physical symmetry} as claimed. This overshadows the physical
viability of formulations that take the operator ${\cal P}$ or
${\cal PT}$ as the fundamental building block
\cite{tanaka,tanaka2} as opposed to the Hilbert space structure
\cite{jpa-2004b,cjp-2004b}.

In general, in order for ${\cal P}$ to describe a
space-reflection, the model(s) under consideration must admit the
$x$ operator as an observable and there must exist a
correspondence principle that relates this operator to the
classical position of a particle. Before specifying an underlying
classical system and formulating a quantization scheme that maps
the classical observables to their quantum analogs, one cannot
assign any physical meaning to the latter. This is the fundamental
reason why one needs a {\em correspondence principle} in any
quantum theory.

The surprising spectral properties of ${\cal PT}$-symmetric
potentials have initially made the impression that these
potentials may be a basis for an extension of quantum mechanics
\cite{bbj}. This turned out not to be true in the sense that such
a theory, even if it is formulated in a consistent manner, will be
equivalent to the conventional quantum mechanics
\cite{critique,cjp-2004b}. What is true is that the formulation
based on ${\cal PT}$-symmetric or more generally pseudo-Hermitian
Hamiltonians yields an alternative (pseudo-Hermitian)
representation of the conventional quantum mechanics. The
availability of this representation seems to have been overlooked
for over seventy years. Its recent discovery
\cite{jpa-2003,jpa-2004b} seems to offer a new prospective for
dealing with a number of important problems of
mathematical/theoretical physics.

A concrete example of the application of the pseudo-Hermitian
representation of quantum mechanics is the solution of the
Hilbert-space problem for Klein-Gordon fields in relativistic
quantum mechanics and Wheeler-DeWitt fields in minisuperspace
quantum cosmology \cite{KG}. Another more recent example is the
treatment of position-dependent-mass Hamiltonians reported in
\cite{bcr}. It is based on the observation that some complex
${\cal PT}$-symmetric potentials define systems that are
perturbatively equivalent to those described by a standard
Hermitian Hamiltonian with a position-dependent mass
\cite{other-models}.

The fact that systems defined by ${\cal PT}$-symmetric potentials
admit a Hermitian description \cite{jpa-2003} is not only
instrumental in determining the observables and the classical
limit of these systems \cite{jpa-2004b}, but it plays an important
role in understanding the spectral properties of the wrong-sign
quartic potential \cite{jm}. Another, perhaps more important,
application of the equivalence of the Hermitian and
pseudo-Hermitian representations of quantum mechanics is a novel
resolution of the old and practically important problem of bound
state scattering \cite{matzkin}.

\end{document}